\documentclass[english,aps,amsmath,amssymb,preprintnumbers]{revtex4}

\usepackage[T1]{fontenc}
\usepackage[latin9]{inputenc}
\usepackage{graphicx}

\makeatletter

\providecommand{\tabularnewline}{\\}

\usepackage{dcolumn}

\usepackage{bm}

\def\order#1{{\cal O}\!\left(#1\right)}
\newcommand{\ep}{\epsilon}
\newcommand{\Li}{{\rm Li}}

\usepackage{babel}
\makeatother

\begin{document}

\preprint{Alberta Thy 10-08}

\title{Heavy-to-heavy quark decays at NNLO}

\author{Alexey Pak and Andrzej Czarnecki}

\affiliation{Department of Physics, University of Alberta, Edmonton, AB, T6G 2G7,
Canada}

\begin{abstract}
Details of a recent calculation of $\order{\alpha_{s}^{2}}$ corrections
to the decay $b\to c\ell\nu_{l}$, taking into account the $c$-quark
mass, are described. Construction of the expansion in the mass ratio
$m_{c}/m_{b}$ as well as the evaluation of new four-loop master integrals
are presented. The same techniques are applicable to the muon decay,
$\mu\to e\nu_{\mu}\bar{\nu}_{e}$. Analytical results are presented,
for the physical cases as well as for a model with purely-vector couplings. 
\end{abstract}
\maketitle

\section{Introduction}

Because of the role that the semileptonic decay $b\to c\ell\nu_{l}$
and the muon decay $\mu\to e\nu_{\mu}\bar{\nu}_{e}$ play in the determination
of the parameters of the Standard Model, it is warranted to improve
the theoretical description of their rates and distributions. Recently,
we have determined $\order{\alpha_{s}^{2}}$ corrections ($\alpha^{2}$
for the muon), including the effects of the charm quark (electron)
mass \cite{Pak:2008qt}. Since this was the first analytical evaluation
of such mass effects and required an extension of known results, in
this paper we describe some technical details. 

We made extensive use of earlier results obtained in the massless
case \cite{vanRitbergen:1998yd,vanRitbergen:1999fi,vanRitbergen:1999gs},
in particular of the master integrals determined in those projects.
It turned out, however, that the massive case requires the knowledge
of more terms of the expansion of those integrals in the dimensional
regularization parameter $\epsilon=\frac{4-D}{2}$. To find them,
we took a different approach to the evaluation of those master integrals.
That approach is described below, together with the list of new terms
that are now known. Since the decays considered here are a model for
other beta-decay-like processes, those integrals will likely find
other applications in the future. 

Before presenting that calculation, in the following Section we introduce
the notation and in Section \ref{sec:asym} describe how the expansion
around the massless case is constructed. Our results for the corrections to the $b\to c $ decay and to a decay in a model with pure vector couplings, as well as for the new terms in master integrals, are collected in three appendices.

\section{Decay rate and radiative corrections}

\label{sec:calc}

\begin{figure}[t]

\begin{centering}
\includegraphics[width=0.24\textwidth]{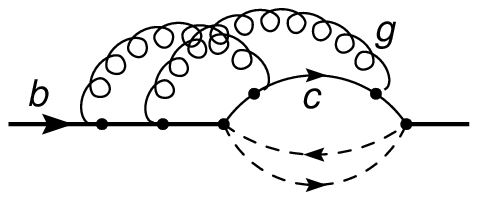} \includegraphics[width=0.24\textwidth]{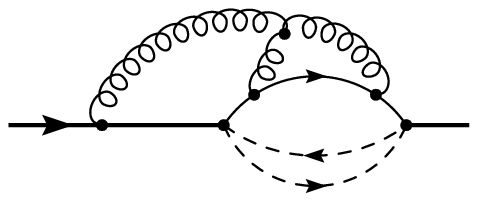}
\includegraphics[width=0.24\textwidth]{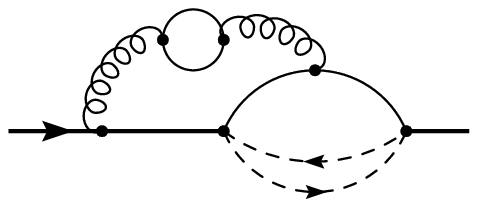} \includegraphics[width=0.24\textwidth]{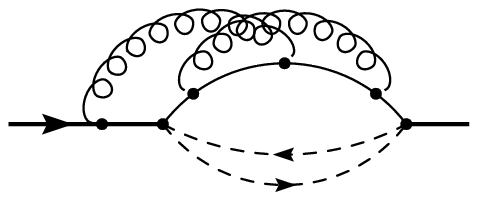}
\\
\parbox[t][1\totalheight]{0.99\textwidth}{%
\caption{\label{fig:dia} Examples of $\order{\alpha_{s}^{2}}$ $b$-quark self-energy
diagrams whose cuts describe the semi-leptonic decay. Dashed lines represent a charged lepton and a neutrino, whose masses we neglect.}
} 
\par\end{centering}
\end{figure}

The tree-level decay rate and the one-loop corrections are known exactly~\cite{Nir:1989rm}.
Their expansion in $\rho\equiv m_{c}/m_{b}\ll1$ and parameterization
of two loop corrections are \begin{eqnarray}
\Gamma(b\to c\ell\bar{\nu}) & = & \Gamma_{0}\left[X_{0}+C_{F}\frac{\alpha_{s}(m_{b})}{\pi}X_{1}+C_{F}\left(\frac{\alpha_{s}(m_{b})}{\pi}\right)^{2}X_{2}+\ldots\right],\label{eqn:gamma}\\
X_{0} & = & 1-8\rho^{2}-24\rho^{4}\ln\rho+8\rho^{6}-\rho^{8},\label{eqn:x0}\\
X_{1} & = & \frac{25}{8}-\frac{\pi^{2}}{2}-\left(34+24\ln\rho\right)\rho^{2}+16\pi^{2}\rho^{3}\label{eqn:x1}\\
 &  & -\left(\frac{273}{2}-36\ln\rho+72\ln^{2}\rho+8\pi^{2}\right)\rho^{4}+16\pi^{2}\rho^{5}-\left(\frac{526}{9}-\frac{152}{3}\ln\rho\right)\rho^{6}+\ldots,\nonumber \\
X_{2} & = & T_{R}N_{L}X_{L}+T_{R}N_{H}X_{H}+T_{R}N_{C}X_{C}+C_{F}X_{A}+C_{A}X_{NA}.\label{eq:x2}\end{eqnarray}
 where $\Gamma_{0}=G_{F}^{2}|V_{cb}|^{2}m_{b}^{5}/\left(192\pi^{3}\right)$,
$G_{F}$ is Fermi constant, and color factors are $C_{F}=\frac{4}{3}$,
$T_{R}=\frac{1}{2}$, $C_{A}=3$. Here $N_{L}=3$ is the number of
massless quarks, and $N_{H,C}=1$ label, respectively, the $b$-quark
loop, and virtual and additional real $c$-quark contributions; components
$X_{L}$, $X_{H}$, $X_{C}$, $X_{A}$, and $X_{NA}$ are separately
gauge-invariant and finite. The limit of these functions for $m_{c}=0$
is known \cite{vanRitbergen:1999fi}. The purpose of this paper is
to obtain their mass dependence. All calculations are performed in
$D=4-2\ep$ dimensions, and axial currents are treated according to
the prescription of \cite{Larin:1993tq}. General gluon gauge was
used to ensure gauge invariance of the results.

The presence of the logarithm of the mass ratio in the lowest-order
rate, Eq.~(\ref{eqn:x0}), signals the presence of higher powers of
logarithms in the higher-order terms (indeed, we see a quadratic logarithm
in Eq.~(\ref{eqn:x1})). It is for this reason that the term $\rho^{4}$
requires more terms of the expansion of Feynman integrals in powers
of $\epsilon$ mentioned in the Introduction. 

Using the optical theorem, we find $X_{2}$ as the imaginary part
of 39 self-energy diagrams such as the examples shown in Fig.~\ref{fig:dia}. Each diagram
is expanded in asymptotic regions \cite{Tkachov:1997gz,Smirnov:2002pj,Czarnecki:1996nr}
to several orders in $\rho$ and all contributions are summed.
Some details of that expansion are described in the next section.

\section{Asymptotic expansion}

\label{sec:asym}

\begin{figure}[t]

\begin{centering}
\includegraphics[width=0.23\textwidth]{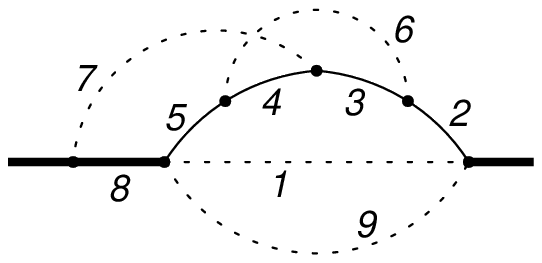} \put(-90,-5){(a)}\hfill{}\includegraphics[width=0.23\textwidth]{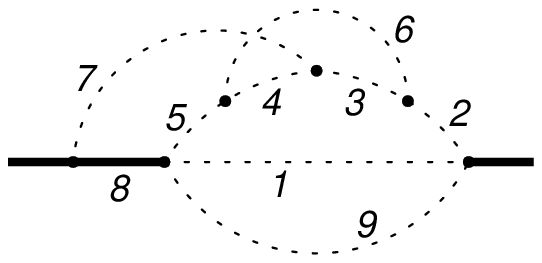}\put(-90,-5){(b)}\hfill{}\includegraphics[width=0.23\textwidth]{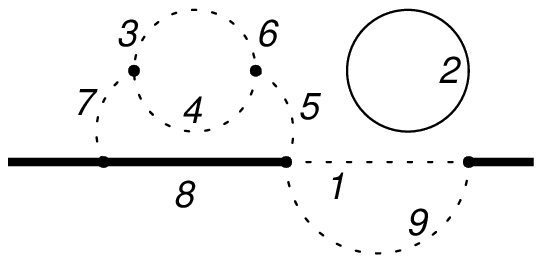}\put(-90,-5){(c)}\hfill{}\includegraphics[width=0.23\textwidth]{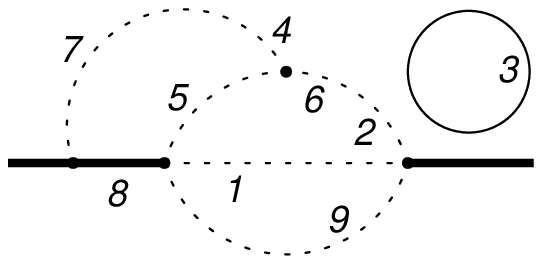}\put(-90,-5){(d)}
\\
 \vspace{0.3cm}
 \includegraphics[width=0.23\textwidth]{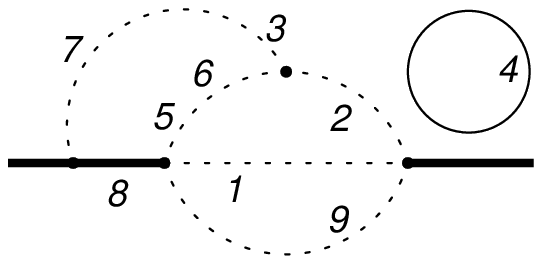}\put(-90,-5){(e)}\hfill{}\includegraphics[width=0.23\textwidth]{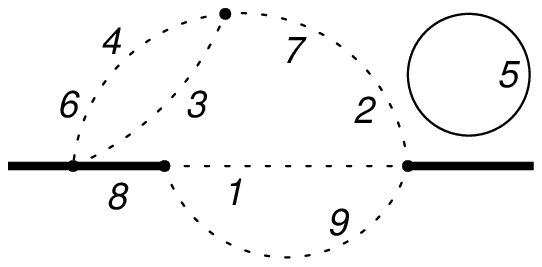}\put(-90,-5){(f)}\hfill{}\includegraphics[width=0.23\textwidth]{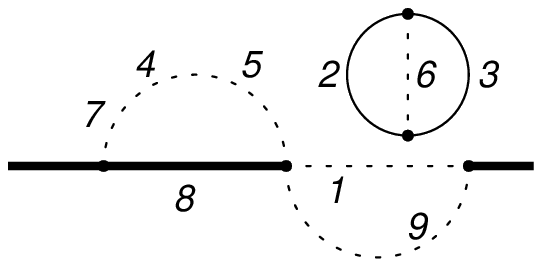}\put(-90,-5){(g)}\hfill{}\includegraphics[width=0.23\textwidth]{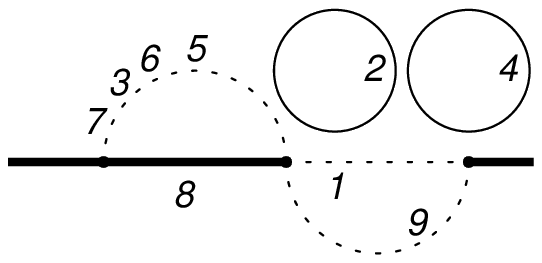}\put(-90,-5){(h)}
\\
 \vspace{0.3cm}
 \includegraphics[width=0.23\textwidth]{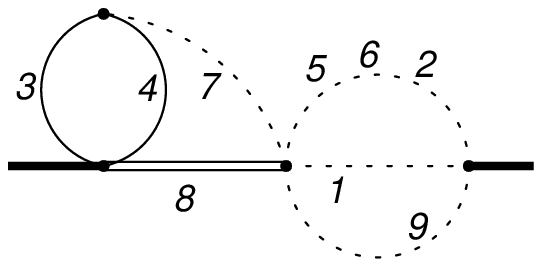}\put(-90,-5){(i)}\hfill{}\includegraphics[width=0.23\textwidth]{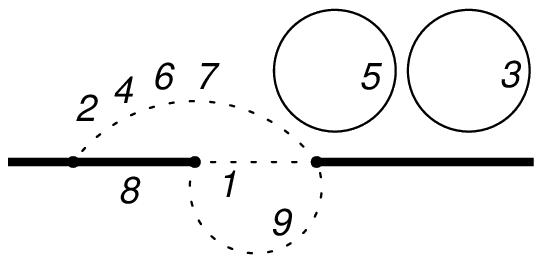}\put(-90,-5){(j)}\hfill{}\includegraphics[width=0.23\textwidth]{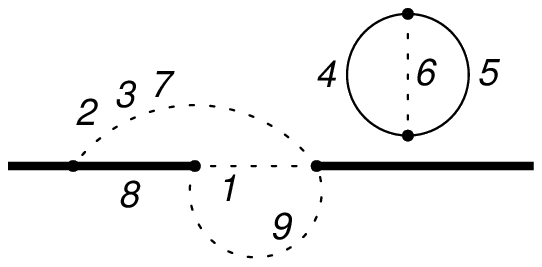}\put(-90,-5){(k)}\hfill{}\includegraphics[width=0.23\textwidth]{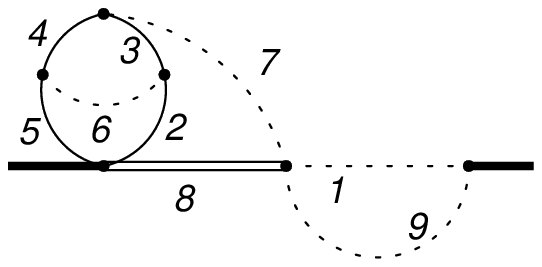}\put(-90,-5){(l)}
\\
 
\par\end{centering}

\caption{\label{fig:regs}Expansion of a double-scale topology (a) in all contributing
asymptotic regions (b-l). Thick lines represent mass $m_{b}$, thin
-- mass $m_{c}$, dashed lines are massless, double lines correspond
to eikonal (static) propagators. Different regions correspond to expansions
with different assignment of scales ({}``hard'', $m_{b}$ or {}``soft'',
$m_{c}$) to the four loop momenta.}

\end{figure}

As an example, we consider the rightmost diagram in Fig.~\ref{fig:dia}
which has the richest structure of asymptotic regions. Evaluating the traces
and Wick rotating, we obtain a number of terms with the following
factors in denominators: \begin{eqnarray}
 &  & D_{1}=k^{2},~D_{2}=(q-r)^{2},~D_{3}=q^{2}+m_{c}^{2},~D_{4}=(q+l)^{2}+m_{c}^{2},\\
 &  & D_{5}=(q+l-r)^{2}+m_{c}^{2}~,D_{6}=r^{2},~D_{7}=l^{2},~D_{8}=l^{2}+2pl,~D_{9}=(p+k-q+r)^{2},\nonumber \end{eqnarray}
where $k$, $q$, $r$, and $l$ are loop momenta, and the external
momentum $p$ is on shell ($p^{2}=-m_{b}^{2}$) (Fig.~\ref{fig:regs}(a)).
Each loop momentum is then assigned one of the two relevant scales
-- $m_{b}$ ({}``hard'') or $m_{c}$ ({}``soft''), and integrands are
expanded in each case. All contributing regions for this topology
are sketched in Fig.~\ref{fig:regs}(b--l). As an example, consider
$q$ and $l$ being {}``soft'', and $k$ and $r$ {}``hard''.
The denominators are Taylor expanded as \begin{eqnarray}
 &  & \frac{1}{D_{8}}=\frac{1}{l^{2}+2pl}=\frac{1}{2pl}\sum_{i=0}^{\infty}\left(-\frac{l^{2}}{2pl}\right)^{i},\mbox{ and similarly }\\
 &  & \frac{1}{D_{2}}\to\sum_{i}\frac{\ldots}{D_{6}^{i}},~~~\frac{1}{D_{5}}\to\sum_{i}\frac{\ldots}{D_{6}^{i}},~~~\frac{1}{D_{9}}\to\sum_{i}\frac{\ldots}{\left[(p+k+r)^{2}\right]^{i}},\nonumber \end{eqnarray}
 producing, together with the remaining denominators, the case (i).

As one can see, in most regions this topology factorizes into known
one- and two-loop integrals. Note the unusual {}``eikonal'' denominator
$1/(2pl+i0)$ in the above example and similar factors in case (l),
which lead to odd powers of $\rho$ in the end result. Disentangling of
the products of loop momenta in the numerator, naturally appearing 
in the expansion, in the most difficult cases involves the algebraic
method of Ref.~\cite{Davydychev:1995nq} and, at high orders in $\rho$, 
leads to a huge number of terms. 
With intermediate expression size reaching 
hundreds of gigabytes, this calculation was possible only with the computer 
algebra system~\cite{Vermaseren:2000nd}.

The three-loop eikonal integrals in region (l) correspond to topologies
studied in Ref.~\cite{Grozin:2006xm}. Using the integration-by-parts
identities~\cite{Tkachev:1981wb,Chetyrkin:1981qh,Laporta:2001dd},
we reduce the expressions to a few {}``master integrals'' found
in Refs.\cite{Grozin:2006xm,Grozin:2007ap}, and a previously unpublished
integral calculated by V. A. Smirnov, Eq.~(\ref{eq:u9}).

The four-loop {}``all-hard'' case (b) here is the most challenging.
To expand $X_{2}$ to $\order{\rho^{7}}$, an implementation of algorithm~\cite{Laporta:2001dd}
was running for several weeks. Finally, all diagrams were reduced
to the 33 master integrals. Their evaluation was the biggest challenge
of this work.

\section{Evaluation of four-loop master integrals}

\label{sec:masters}

The results for master integrals given in
\cite{vanRitbergen:1999fi}
are sufficient to obtain $\order{\rho^{0}}$ and $\order{\rho^{2}}$
terms of $X_{2}$. However, in order to find the following $\order{\rho^{4}}$
contribution, many integrals need to be expanded further.
In \cite{vanRitbergen:1999fi}, the initial terms of that expansion were obtained in a series of steps.  First, some of the internal lines of the diagram, representing a given master integral, were assigned a mass $M$ much larger than the mass of the external particle $m$.  Next, the diagram was expanded in  $m/M$ using a similar approach as described in the previous section; several terms of that expansion were obtained in each order in $\epsilon$.  Finally, and most non-trivially, a pattern in that expansion was recognized and the expansion was now summed analytically.  The value of the result at $M=m$ gave the desired value of the integral. 

\begin{figure}[b]
\parbox[c][1\totalheight]{0.25\textwidth}{%
 \includegraphics[width=0.3\textwidth]{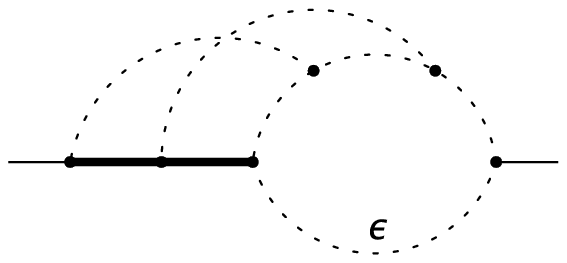}%
} %
\parbox[c][1\totalheight]{0.74\textwidth}{%
 \begin{eqnarray*}
 &  & {I}(a_{1},...,a_{9};x)=\frac{1}{\pi\mathcal{F}^4}{\rm Im}\int\frac{[d^{D}k][d^{D}q][d^{D}r]~~~D_{9}^{-a_{9}}}{D_{1}^{a_{1}+\ep}D_{2}^{a_{2}}D_{3}^{a_{3}}D_{4}^{a_{4}}D_{5}^{a_{5}}D_{6}^{a_{6}}D_{7}^{a_{7}}D_{8}^{a_{8}}},\\
 &  & \begin{array}{l}
p^{2}=-1,D_{1}=k^{2},~D_{2}=(k+p)^{2},~D_{3}=(k+q+p)^{2},\\
D_{4}=(k+q+r+p)^{2},~D_{5}=(q+r+p)^{2}+1/x,\\
D_{6}=(r+p)^{2}+1/x,~D_{7}=q^{2},~D_{8}=r^{2},~D_{9}=2qp.\end{array}\end{eqnarray*}
} \\

\caption{\label{fig:t3be56}Auxiliary double-scale topology. Thick lines represent
mass $\sqrt{1/x}$, thin -- unit mass propagators, and $\ep$ labels
a denominator raised to a non-integer power.}

\end{figure}

In higher orders in $\epsilon$, the recognition of the expansion pattern may be very difficult.  Instead, 
we evaluate these integrals using the method of differential equations~\cite{Remiddi:1997ny}.
We start by choosing a few artificial double-scale topologies, related
in some limit to the needed integrals.  We illustrate this approach with the topology shown
in Fig.~\ref{fig:t3be56}.  It involves an artificial mass $\sqrt{1/x}$,
and in the on-shell limit $x\to1$ reproduces the topology needed for
the first diagram shown in Fig.~\ref{fig:dia} (where we also introduce the 
loop factor $\mathcal{F} = \frac{\Gamma(1 + \ep)}{(4\pi)^{D/2}}$).

This function is chosen due to several reasons. First, if $0<x\le1$,
the cuts are the same as in the on-shell limit, and we may discard
the real part in all calculations. Second, large mass expansion to
a few orders in $x$ and $\ep$ is relatively simple~\cite{vanRitbergen:1999fi}.
And third, the associated differential equations have a structure convenient
for an iterative solution; this property will be explained later.

This topology has 40 master integrals. Derivative of any such integral
can be re-cast in terms of integrals with shifted indices, \begin{equation}
\frac{\partial}{\partial x}I(a_{1},...,a_{9};x)=\frac{a_{5}}{x^{2}}I(a_{1},...,a_{5}+1,...)+\frac{a_{6}}{x^{2}}I(a_{1},...,a_{6}+1,...),\end{equation}
generating a system of 40 differential equations. It can be split
into independent subsystems of at most four equations, where the solution
of one system enters the RHS of the following one.

For example,  consider functions 
$f(x,\ep)={I}(0,1,1,0,1,0,0,1,0;x)$,
$g(x,\ep)={I}(-1,1,1,0,1,0,0,1,0;x)$, and 
$h(x,\ep)={I}(0,1,1,-1,1,0,0,1,0;x)$,
entering a particularly simple system of relations \begin{eqnarray}
f^{\prime} & = & \frac{4x-5+\ep(7-6x)}{2x(1-x)}f+\frac{9-11\ep}{2(1-x)}g+\frac{3\ep-3}{2(1-x)}h,\\
g^{\prime} & = & \frac{1-\ep}{2x(1-x)}f-\frac{8+x-\ep(10+x)}{2x(1-x)}g+\frac{3-3\ep}{2(1-x)}h,\nonumber \\
h^{\prime} & = & \frac{1-\ep}{2x^{2}}f+\frac{1-\ep}{2x}g+\frac{3\ep-3}{2x}h.\nonumber \end{eqnarray}

We solve this system with respect to derivatives of $f$, and expand
$f(x,\ep)=\frac{1}{\ep}f_{-1}(x)+f_{0}(x)+\ep f_{1}(x)+...$. Finally,
the differential equations for $f_{i}$ become 
\begin{eqnarray}
 &  & f_{i}^{\prime\prime\prime}+\frac{9-6x}{x(1-x)}f_{i}^{\prime\prime}+\frac{18-6x}{x^{2}(1-x)}f_{i}^{\prime}+\frac{6}{x^{3}(1-x)}f_{i}=R_{i},\\
 &  & R_{-1}=0,~~~R_{0}=\frac{10-4x}{x(1-x)}f_{-1}^{\prime\prime}+\frac{47-9x}{x^{2}(1-x)}f_{-1}^{\prime}+\frac{31}{x^{3}(1-x)}f_{-1},~~~\ldots~.\nonumber \end{eqnarray}

The three solutions of the homogeneous equation (with $R_{i}=0$) can
be guessed: $1/x$, $1/x^{2}$, and $(1-x^{2}(x+6)+6x(1+x)\ln x)/x^{3}$.
Euler's formula allows then to solve the inhomogeneous equations. To fix
the three integration constants, we use the large mass expansion \begin{eqnarray}
f(x,\ep) & = & -~\frac{1}{4x\ep}+\frac{1}{24}-\frac{15+4\ln x}{8x}\\
 &  & +~\ep\left(\frac{2\pi^{2}-145-60\ln x-8\ln^{2}x}{16x}+\frac{1}{4}+\frac{\ln x}{12}+\frac{x}{144}+\frac{x^{2}}{1440}+\ldots\right)+\ldots~.\nonumber \end{eqnarray}

The general solution for $f_{i}$ is expanded and matched to this
series to find suitable constants. Finally, we obtain $x$-dependent
solutions: \begin{eqnarray}
f_{-1} &= & -~\frac{1}{4x}, 
\qquad f_{0} = \frac{1}{24}-\frac{15+4\ln x}{8x},
\qquad f_{1} = \frac{19}{36}+\frac{1}{12x^{2}}+\frac{6\pi^{2}-373}{48x}
\\ \nonumber &&
+\frac{H(0;x)(x-45)}{12x}+\frac{H(1;x)(x^{3}+9x^{2}-9x-1)}{12x^{3}}-\frac{H(0,1;x)(1 + x)}{2x^{2}}-\frac{H(0,0;x)}{x},~\ldots~.\nonumber \end{eqnarray}
 Due to the fact that the answers can be expressed in terms of harmonic
polylogarithms (HPLs) $H(...;x)$ \cite{Remiddi:1999ew}, the expansion
can be continued as long as CPU resources allow. (More accurately, this procedure works as long as
the solutions of the homogeneous equation, as well as their inverse Wronskian and
minors of Wronski matrix depend on $x$ in the denominators only through 
factors $x$ and $1\pm x$, and numerators contain HPLs and polynomials.  This allows to solve the inhomogeneous equation in terms of HPLs.) Taking
the limit $x\to1$, we obtain the required on-shell integral. 

One integral of the topology  shown in Fig.~\ref{fig:t3be56},
$u(x,\ep)={I}(0,1,1,1,1,1,1,1,0;x)$, 
presents an additional difficulty.
Its $x$-dependent expression starts at $\order{\ep}$ and logarithmically
diverges at $x=1$. To find that integral, we evaluated instead the
infrared-safe function ${I}(1,1,1,1,1,1,1,1,-1;x)$, took the
on-shell limit and then reduced to $u(1,\ep)$. This lead to finite
$\order{\ep^0}$ and $\order{\ep}$ expressions:
$u(1,\ep) = -~\frac{4\pi^4}{135} - \ep\left(\frac{95\zeta_5}{6} 
  + \frac{13\pi^2\zeta_3}{18} + \frac{8\pi^4}{135}\right)$.
As a tradeoff,
other integrals were needed to a higher order in $\ep$, involving
harmonic polylogarithms up to weight six.

An important tool was the software package HPL \cite{Maitre:2005uu}
and additional software for the series expansion of harmonic polylogarithms
was developed \cite{Pak:HPL}. As an independent check, we used numerical
integration of the Mellin-Barnes representation of some integrals
\cite{Czakon:2005rk}.

\section{Results and summary}

\label{sec:results}

\begin{figure}[h]

\begin{raggedright}
\includegraphics[width=0.45\textwidth]{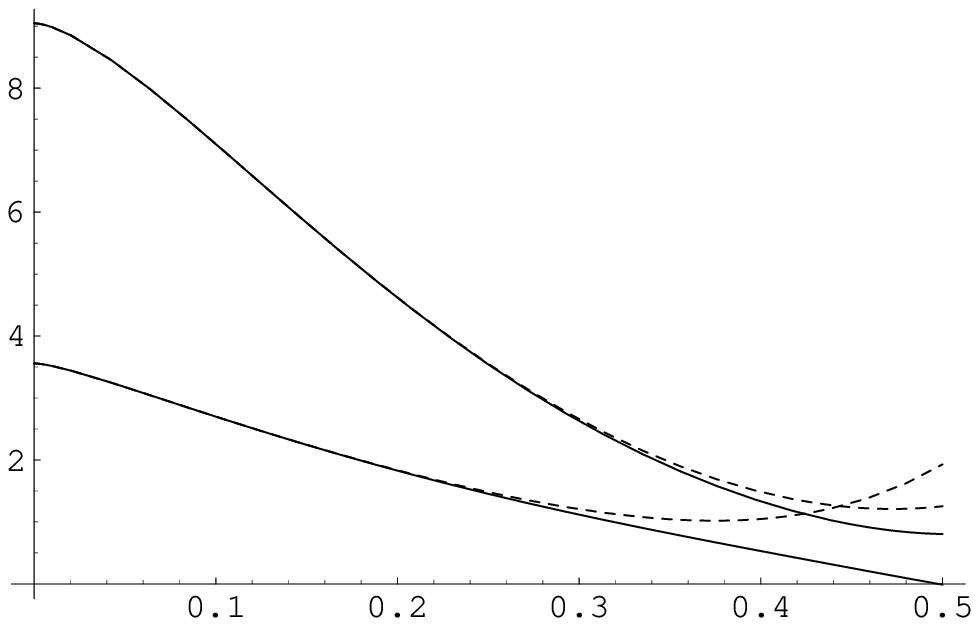} \put(10,5){\mbox{$\rho$}}
\put(-150,50){\mbox{$X_{A}(\rho)$}} \put(-110,70){\mbox{$-X_{NA}(\rho)$}}
\hspace{0.7cm} 
\includegraphics[width=0.45\textwidth]{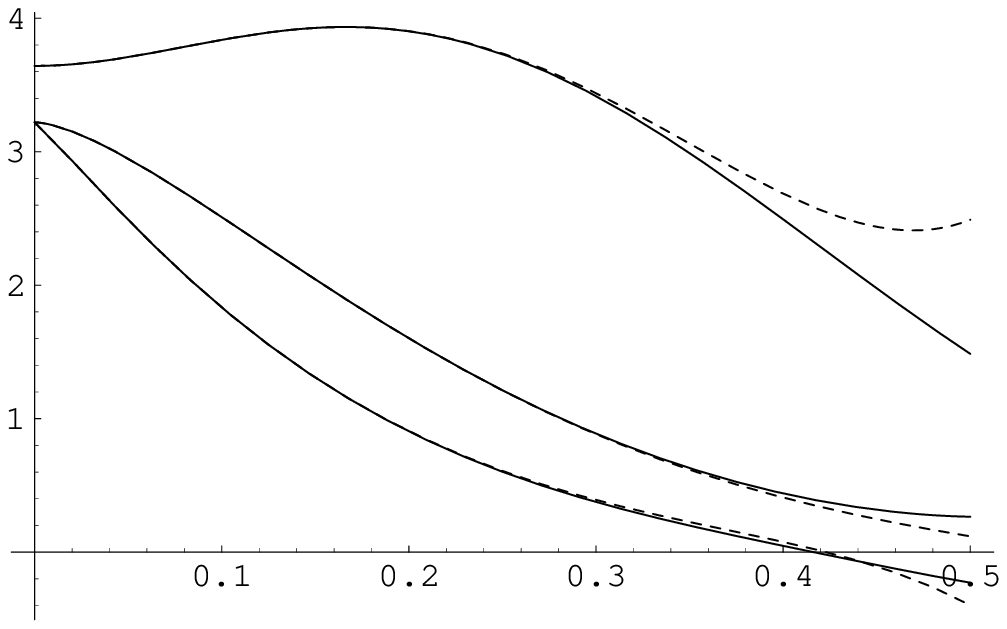} \put(10,5){\mbox{$\rho$}}
\put(-175,35){\mbox{$X_{C}(\rho)$}} \put(-130,70){\mbox{$X_{L}(\rho)$}}
\put(-70,110){\mbox{$-10^{2}~X_{H}(\rho)$}} \hfill{}\\
 
\par\end{raggedright}

\begin{centering}
\parbox[t][1\totalheight]{0.99\textwidth}{%
\caption{\label{fig:plots} Mass-dependent corrections to $X_{2}$ of Eq.~(\ref{eqn:gamma}). }
} 
\par\end{centering}
\end{figure}

The rather lengthy expressions for $X_{L}$, $X_{H}$, $X_{C}$, $X_{A}$,
and $X_{NA}$ calculated through $\order{\rho^{7}}$ are presented
in  Appendix \ref{sec:appendix}.  Their features, such as the logarithms and even and odd powers of $\rho$, together with
physics consequences,
have been discussed in \cite{Pak:2008qt}.    Here we would like to illustrate the convergence of the expansion.  To this end, 
Fig.~\ref{fig:plots} shows the plots of $X_i$ 
as functions of $\rho$, in two versions: solid lines show all known terms while dashed lines are obtained by leaving out the 
last known power of $\rho$.  We see that the convergence is excellent up to at least $\rho=0.3$, of interest in this study.  

\begin{figure}[t]
\begin{raggedright}
\includegraphics[width=0.45\textwidth]{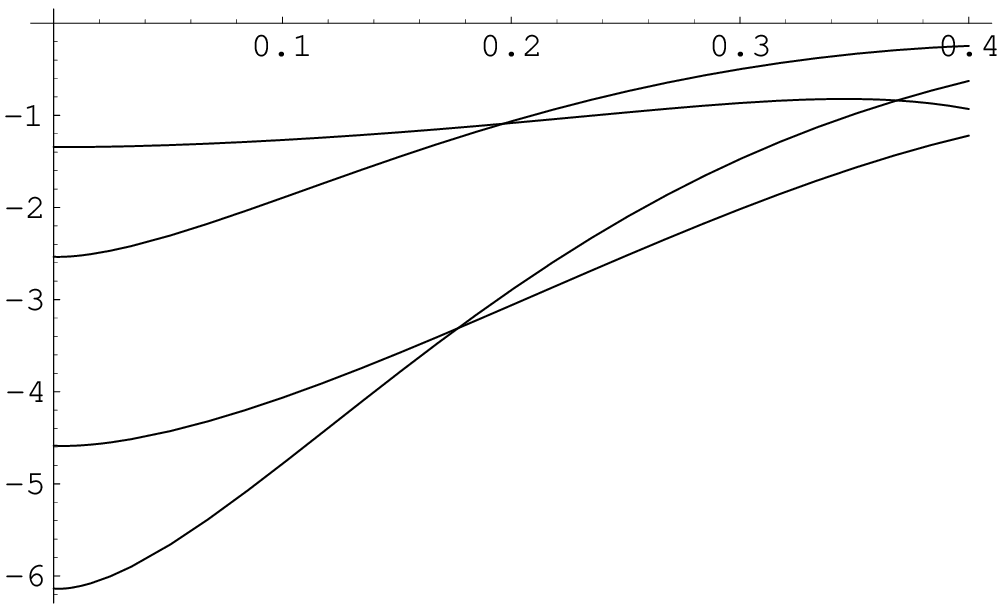} 
\put(10,130){$\rho$}
\put(-160,20){$L_{2}^{(1)}$} 
\put(-40,85){$H_{2}^{(1)}$} 
\put(-180,75){$L_{2}^{(2)}$}
\put(-190,115){$H_{2}^{(2)}$} \hspace{0.7cm} 
\includegraphics[width=0.45\textwidth]{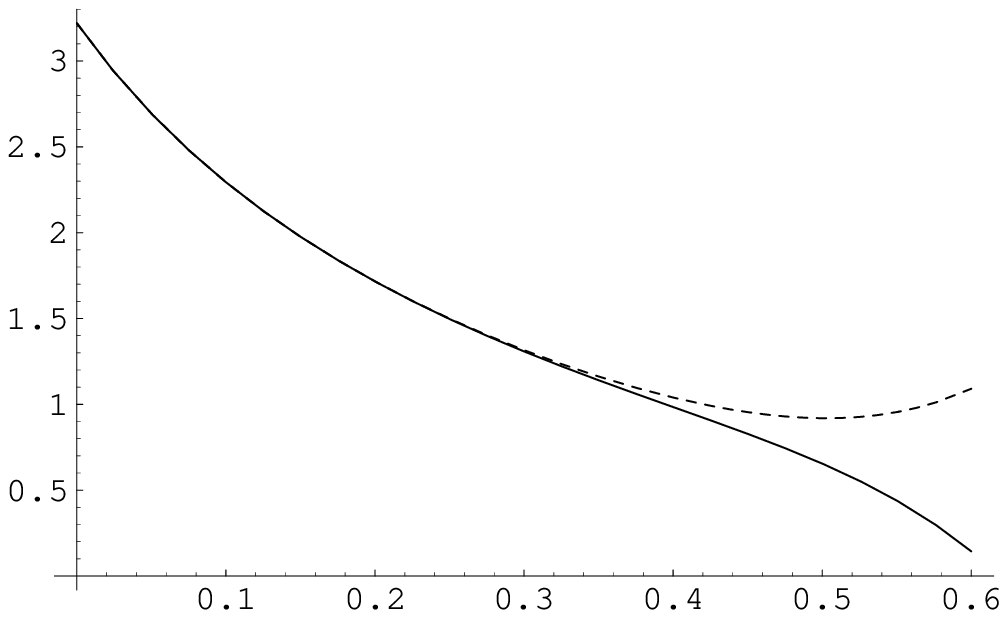}
\put(10,5){$\rho$} \put(-120,80){$U_{C}$} 
\par\end{raggedright}
\begin{centering}
\parbox[t][1\totalheight]{0.45\textwidth}{%
\caption{\label{fig:mom}First two moments of lepton and hadron energy distributions}
}\hfill{}%
\parbox[t][1\totalheight]{0.45\textwidth}{%
\caption{\label{fig:uc}Charm quark contribution to semi-leptonic $b\to u$
decays}
} 
\par\end{centering}
\end{figure}

In addition to the integrated decay rate, precision fits to experimental
data are done for the moments of the lepton energy ${E}_{l}$ and the 
hadronic-system energy $E_{h}$ distributions in the rest frame of the $b$-quark,
with the goal of accurately measuring several parameters including $|V_{cb}|$,
$m_{b}$, and Wilson coefficient of non-perturbative operators. 
Thus, QCD corrections to those moments are also of interest.
These 
corrections  are defined by
 \begin{equation}
\int\left(\frac{E_{l}}{m_{b}}\right)^{n}d\Gamma=\Gamma_{0}\left[L_{0}^{(n)}+C_{F}\frac{\alpha_{s}}{\pi}L_{1}^{(n)}+C_{F}\left(\frac{\alpha_{s}}{\pi}\right)^{2}L_{2}^{(n)}\right],\end{equation}
 and similarly for the moments of $E_{h}$, described by coefficients
$H_{j}^{(n)}$; the average is taken over the whole phase space of
decay products. One-loop spectra are known exactly~\cite{Czarnecki:1990pe},
and Fig.~\ref{fig:mom} presents NNLO corrections $L_{2}^{(1,2)}$
and $H_{2}^{(1,2)}$. 
In \cite{Pak:2008qt} we discussed how the experimental cuts can be approximately modeled by the analytical calculation.
Finally, we present the analytical results for $U_{C}$ (Eq.~(\ref{eqn:uc}) and
Fig.~\ref{fig:uc}), an analogue of $X_{C}$ in Eq.~(\ref{eqn:gamma}),
describing charm quark loop contribution to the process with a massless quark in the final state, 
$b\to u\ell\bar{\nu}$.

The same method can be used in the model in which
chiral weak coupling of quarks, $\frac{ig_{w}}{2\sqrt{2}}\gamma_{\mu}\left(1-\gamma_{5}\right)$,
are replaced with a pure vector vertex, $\frac{ig_{w}}{2}\gamma_{\mu}$. This is
a useful toy model, e.g., for logarithmic re-summation studies~\cite{Bauer:1996ma}.
In parameterization similar to Eq.~(\ref{eqn:gamma}), the second order
components $V_{i}$ are defined in Eqs.~(\ref{eqn:vn}--\ref{eqn:vc}).
We evaluated terms to $\order{\rho^{5}}$, and it is now easy to find
more.

To summarize, in the process of this calculation, we checked and confirmed
the massless limit of the $\order{\alpha_{s}^{2}}$ corrections 
\cite{vanRitbergen:1998yd,vanRitbergen:1999fi,vanRitbergen:1999gs}.
We extended those results to several orders in the mass ratio of the
final and initial quarks. To complete this task, additional terms
of master integrals were required, and the differential-equation method
was applied to compute them. Our results agree excellently with the
numerical calculation \cite{Melnikov:2008qs}. An ultimate test of
the convergence of the mass expansion will be a corresponding expansion
around the opposite mass limit, $m_{c}\simeq m_{b}$. Work on this
is in progress.

\begin{acknowledgments}
We thank A. Davydychev, M. Dowling, M. Kalmykov, A. Kotikov, A. Penin, and J. Piclum
for many helpful discussions. This work was supported by Science and
Engineering Research Canada.
\end{acknowledgments}
\appendix

\section{Analytical results}

\label{sec:appendix}

Here we present the components of the total decay rate evaluated to
$\order{\rho^{7}}$. In the case of the $b$-quark loop contribution,
$X_{H}$, Fig. \ref{fig:plots} shows an extremum point around $\rho=0.2$.
This contrasts with the other contributions, that seem to be monotonous functions of $\rho$. 
In order to double-check the convergence, also the $\order{\rho^{8}}$
term has been computed. The final results are 
\begin{eqnarray}
X_{L} & = & -~\frac{1009}{288}+\frac{8\zeta_{3}}{3}+\frac{77\pi^{2}}{216}+\left\{ \frac{118}{3}-\frac{4\pi^{2}}{3}+\frac{52}{3}\ln\rho-8\ln^{2}\rho\right\} \rho^{2}\label{eqn:xl}\\
 & + & \left\{ \frac{64\ln2}{3}-\frac{112}{9}+\frac{32}{3}\ln\rho\right\} \pi^{2}\rho^{3}+\left\{ 76\zeta_{3}-\frac{5\pi^{2}}{3}-33+52\ln^{2}\rho\right.\nonumber \\
 &  & \left.+\left(39-\frac{16}{3}\pi^{2}\right)\ln\rho-32\ln^{3}\rho\right\} \rho^{4}+\left\{ \frac{64\ln2}{3}-\frac{1216}{45}+\frac{32}{3}\ln\rho\right\} \pi^{2}\rho^{5}\nonumber \\
 & + & \left\{ \frac{344}{27}+\frac{28\pi^{2}}{27}-\frac{1564}{27}\ln\rho+24\ln^{2}\rho\right\} \rho^{6}+\frac{40}{21}\pi^{2}\rho^{7},\nonumber \\
X_{C} & = & -~\frac{1009}{288}+\frac{8\zeta_{3}}{3}+\frac{77\pi^{2}}{216}-\frac{5}{4}\pi^{2}\rho+\left\{ \frac{145}{3}+\frac{16\pi^{2}}{3}+\frac{52}{3}\ln\rho-8\ln^{2}\rho\right\} \rho^{2}\label{eqn:xc}\\
 & + & \left\{ \frac{569}{36}+\frac{64}{3}\ln\rho\right\} \pi^{2}\rho^{3}+\left\{ 196\zeta_{3}+\frac{\pi^{2}}{6}-\frac{4483}{36}+44\ln^{2}\rho-32\ln^{3}\rho\right.\nonumber \\
 &  & \left.+\left(\frac{599}{6}+\frac{74\pi^{2}}{3}\right)\ln\rho\right\} \rho^{4}+\left\{ \frac{50}{3}\ln\rho-\frac{172}{9}\right\} \pi^{2}\rho^{5}\nonumber \\
 & - & \left\{ \frac{33982}{225}+\frac{232\pi^{2}}{27}-\frac{11836}{135}\ln\rho+\frac{64}{9}\ln^{2}\rho\right\} \rho^{6}+\left\{ \frac{44}{3}+18\ln\rho\right\} \pi^{2}\rho^{7},\nonumber \\
X_{H} & = & \frac{16987}{576}-\frac{64\zeta_{3}}{3}-\frac{85\pi^{2}}{216}+\left\{ \frac{8\pi^{2}}{3}-\frac{1198}{45}\right\} \rho^{2}+\left\{ \frac{156901877}{2116800}-\frac{11\pi^{2}}{18}\right.\label{eqn:xh}\\
 &  & \left.-~64\zeta_{3}-\left(\frac{186689}{2520}-\frac{20\pi^{2}}{3}\right)\ln\rho\right\} \rho^{4}+\left\{ \frac{189825233}{7144200}-\frac{52\pi^{2}}{27}-\frac{181627}{14175}\ln\rho\right.\nonumber \\
 &  & \left.+~\frac{16}{5}\ln^{2}\rho\right\} \rho^{6}+\left\{ \frac{629309}{1403325}-\frac{4\zeta_{3}}{3}+\frac{19\pi^{2}}{72}-\left(\frac{4741}{9072}+\frac{\pi^{2}}{9}\right)\ln\rho\right\} \rho^{8},\nonumber \\
X_{A} & = & \frac{11047}{2592}-\frac{223\zeta_{3}}{36}-\frac{515\pi^{2}}{81}+\frac{53\pi^{2}\ln2}{6}+\frac{67\pi^{4}}{720}+\left\{ \frac{497\pi^{2}}{108}-\frac{2089}{8}+86\zeta_{3}\right.\label{eqn:xa}\\
 &  & \left.-~8\pi^{2}\ln2+\frac{121\pi^{4}}{540}-105\ln\rho-36\ln^{2}\rho\right\} \rho^{2}+\left\{ \frac{752}{9}-\frac{112\pi}{3}\right\} \pi^{2}\rho^{3}\nonumber \\
 & + & \left\{ \frac{16586}{27}-\frac{1139\pi^{2}}{24}-\frac{795\zeta_{3}}{2}+\frac{415\pi^{2}\zeta_{3}}{3}+13\pi^{2}\ln2-96~\Li_{4}\frac{1}{2}-8\pi^{2}\ln^{2}2\right.\nonumber \\
 &  & \left.-~4\ln^{4}2-144\ln^{3}\rho-\left(\frac{19459}{18}+\frac{71\pi^{2}}{3}-246\zeta_{3}+60\pi^{2}\ln2-\frac{40\pi^{4}}{3}\right)\ln\rho\right.\nonumber \\
 &  & \left.-~\frac{349\pi^{4}}{72}+\left(99+4\pi^{2}\right)\ln^{2}\rho+935\zeta_{5}\right\} \rho^{4}+\left\{ \frac{67448}{675}-\frac{776\pi}{15}+160\ln\rho\right\} \pi^{2}\rho^{5}\nonumber \\
 & + & \left\{ \frac{4859\zeta_{3}}{12}-\frac{1732}{9}-\frac{14921\pi^{2}}{216}+\frac{89\pi^{2}\ln2}{6}-\frac{10\pi^{4}}{3}+\left(\frac{1862}{9}-\frac{34\pi^{2}}{3}\right)\ln^{2}\rho\right.\nonumber \\
 &  & \left.-\left(\frac{3635}{18}+136\zeta_{3}-\frac{833\pi^{2}}{18}\right)\ln\rho\right\} \rho^{6}+\left\{ \frac{86\pi}{7}-\frac{469304}{11025}+\frac{1376}{45}\ln\rho\right\} \pi^{2}\rho^{7},\nonumber \end{eqnarray}

\begin{eqnarray}
X_{NA} & = & -\frac{X_{A}}{2}+\frac{19669}{1152}-\frac{70\pi^{2}}{27}+\frac{7\pi^{4}}{60}-\frac{101\zeta_{3}}{12}+\left\{ \frac{11\pi^{4}}{18}-\frac{1813}{8}-\frac{19\pi^{2}}{6}-\frac{685}{6}\ln\rho+4\ln^{2}\rho\right\} \rho^{2}\label{eq:xn}\\
 &  & +\left\{ \frac{2044}{9}-\frac{1136\ln{2}}{3}-\frac{124}{3}\ln\rho\right\} \pi^{2}\rho^{3}+\left\{ \frac{8947}{32}-\frac{103\pi^{2}}{12}+\frac{2\pi^{4}}{45}+200\zeta_{5}-\frac{705\zeta_{3}}{2}+\frac{100\pi^{2}\zeta_{3}}{3}\right.\nonumber \\
 &  & \quad\left.+\left(84\zeta_{3}-\frac{1049}{2}-\frac{161\pi^{2}}{6}+\frac{10\pi^{4}}{3}\right)\ln\rho+\left(7\pi^{2}-\frac{271}{2}\right)\ln^{2}\rho+16\ln^{3}\rho\right\} \rho^{4}\nonumber \\
 &  & +\left\{ \frac{58024}{225}-\frac{1136\ln{2}}{3}-\frac{292}{15}\ln\rho\right\} \rho^{5}\pi^{2}+\left\{ \frac{269297}{1296}+\frac{2303\pi^{2}}{216}-\frac{\pi^{4}}{2}-24\zeta_{3}\right.\nonumber \\
 &  & \quad\left.+\left(12\zeta_{3}-\frac{2441}{72}-\frac{17\pi^{2}}{3}\right)\ln\rho+\left(\frac{229}{9}+\pi^{2}\right)\ln^{2}\rho\right\} \rho^{6}-\left\{ \frac{242554}{33075}+\frac{256}{315}\ln\rho\right\} \pi^{2}\rho^{7}.\nonumber \end{eqnarray}

Charm quark contribution to $b\to u\ell\bar{\nu}$ decays through
$\order{\rho^{7}}$ is: \begin{eqnarray}
U_{C} & = & -~\frac{1009}{288}+\frac{8\zeta_{3}}{3}+\frac{77\pi^{2}}{216}-\frac{5}{4}\pi^{2}\rho+\left\{ 21+\frac{8\pi^{2}}{3}\right\} \rho^{2}+\left\{ \frac{64\ln2}{3}-\frac{95}{36}+\frac{32}{3}\ln\rho\right\} \pi^{2}\rho^{3}\label{eqn:uc}\\
 &  & +\left\{ 48\zeta_{3}-\frac{4375}{36}-\frac{25\pi^{2}}{6}+\left(\frac{365}{6}+6\pi^{2}\right)\ln\rho-8\ln^{2}\rho\right\} \rho^{4}-\frac{112}{15}\pi^{2}\rho^{5}\nonumber \\
 &  & +\left\{ \frac{7804}{675}+\frac{64\pi^{2}}{27}+\frac{8}{5}\ln\rho-\frac{64}{9}\ln^{2}\rho\right\} \rho^{6}-\frac{24}{7}\pi^{2}\rho^{7}.\nonumber \end{eqnarray}

\section{A vector model}
\label{sec:vector}
As a byproduct of this project, for the purpose of comparisons with
\cite{Bauer:1996ma}, we have determined $\order{\alpha_{s}^{2}}$
corrections to the decay rate in a model where the $W$-boson has
only a vector, and no axial-vector, coupling to fermions. That is,
its interaction vertex is obtained from the standard ($V-A$) one
by the substitution $\frac{ig_{w}}{2\sqrt{2}}\gamma_{\mu}\left(1-\gamma_{5}\right)\to\frac{ig_{w}}{2}\gamma_{\mu}$.
The results for this model are parameterized in analogy with Eqs.
(\ref{eqn:gamma},\ref{eqn:x0},\ref{eqn:x1},\ref{eq:x2}), with
replacements $X_{i}\to V_{i}$. Results for the tree-level and first-order
corrections, expanded through $\order{\rho^{5}}$, are \begin{eqnarray}
V_{0} & = & 1-2\rho-8\rho^{2}-\left\{ 18+24\ln\rho\right\} \rho^{3}-24\rho^{4}\ln\rho+\left\{ 18-24\ln\rho\right\} \rho^{5},\label{eqn:vn}\\
V_{1} & = & \frac{25}{8}-\frac{\pi^{2}}{2}+\left\{ \pi^{2}-10-3\ln\rho\right\} \rho-\left\{ 34+24\ln\rho\right\} \rho^{2}+\left\{ 13\pi^{2}-90-81\ln\rho-36\ln^{2}\rho\right\} \rho^{3}\\
 &  & +\left\{ 24\pi^{2}-\frac{273}{2}+36\ln\rho-72\ln^{2}\rho\right\} \rho^{4}+\left\{ 13\pi^{2}-\frac{369}{2}+135\ln\rho-72\ln^{2}\rho\right\} \rho^{5}.\end{eqnarray}

The difference with the standard decay is seen already at the tree
level: odd powers of $\rho$ are present in $V_{0}$, while in the
$V-A$ decay they appeared only in the first-order corrections. The
reason for this is a contribution of a four-quark operator, as discussed
in \cite{Bauer:1996ma,Pak:2008qt}. In the $V-A$ decay, a similar
operator contributes only through its $\order{\alpha_{s}}$ matrix
element. As a result, the second-order corrections in the vector case
are even more complicated than in the $V-A$ decay. They read

\begin{eqnarray}
V_{A} & = & \frac{11047}{2592}-\frac{223\zeta_{3}}{36}-\frac{515\pi^{2}}{81}+\frac{53\pi^{2}\ln2}{6}+\frac{67\pi^{4}}{720}\label{eqn:va}\\
 & + & \left\{ \frac{1997\pi^{2}}{648}-\frac{81149}{1944}+\frac{317\zeta_{3}}{18}-\frac{\pi^{2}\ln2}{3}-\frac{211\pi^{4}}{1080}-\frac{9}{4}\ln^{2}\rho+\left(\frac{3\pi^{2}}{2}-\frac{123}{8}\right)\ln\rho\right\} \rho\nonumber \\
 & + & \left\{ \frac{497\pi^{2}}{108}-\frac{2089}{8}+86\zeta_{3}-8\pi^{2}\ln2+\frac{121\pi^{4}}{540}-105\ln\rho-36\ln^{2}\rho\right\} \rho^{2}\nonumber \\
 & + & \left\{ \frac{4667\pi^{2}}{72}-\frac{16943}{36}+561\zeta_{5}-\frac{37\zeta_{3}}{2}-48~\Li_{4}\frac{1}{2}-2\ln^{4}2+83\pi^{2}\zeta_{3}-17\pi^{2}\ln2\right.\nonumber \\
 &  & \left.+\left(8\pi^{4}+328\zeta_{3}+\frac{125\pi^{2}}{6}-\frac{50969}{72}-64\pi^{2}\ln2\right)\ln\rho-\frac{112\pi^{3}}{3}-8\pi^{2}\ln^{2}2+\frac{449\pi^{4}}{120}\right.\nonumber \\
 &  & \left.+\left(12\pi^{2}-\frac{747}{4}\right)\ln^{2}\rho-63\ln^{3}\rho\right\} \rho^{3}+\left\{ \frac{16586}{27}+\frac{13223\pi^{2}}{72}-96~\Li_{4}\frac{1}{2}+13\pi^{2}\ln2\right.\nonumber \\
 &  & \left.-~\frac{795\zeta_{3}}{2}+\frac{415\pi^{2}\zeta_{3}}{3}+935\zeta_{5}-4\ln^{4}2-8\pi^{2}\ln^{2}2-\frac{304\pi^{3}}{3}-\frac{349\pi^{4}}{72}-144\ln^{3}\rho\right.\nonumber \\
 &  & \left.+\left(99+4\pi^{2}\right)\ln^{2}\rho+\left(\frac{40\pi^{4}}{3}-\frac{19459}{18}+246\zeta_{3}+\frac{361\pi^{2}}{3}-60\pi^{2}\ln2\right)\ln\rho\right\} \rho^{4}\nonumber \\
 & + & \left\{ \frac{14036}{27}+561\zeta_{5}+\frac{109\zeta_{3}}{2}-24~\Li_{4}\frac{1}{2}-\ln^{4}2+\frac{264109\pi^{2}}{5400}+83\pi^{2}\zeta_{3}+41\pi^{2}\ln2\right.\nonumber \\
 &  & \left.+\left(81\zeta_{3}-\frac{8456}{9}+\frac{401\pi^{2}}{2}-66\pi^{2}\ln2+8\pi^{4}\right)\ln\rho+\left(\frac{1395}{4}-8\pi^{2}\right)\ln^{2}\rho\right.\nonumber \\
 &  & \left.-~141\ln^{3}\rho-8\pi^{2}\ln^{2}2-\frac{776\pi^{3}}{15}-\frac{1093\pi^{4}}{90}\right\} \rho^{5},\nonumber 
\\
V_{NA} & = & -\frac{V_{A}}{2}+\frac{19669}{1152}-\frac{70\pi^{2}}{27}+\frac{7\pi^{4}}{60}-\frac{101\zeta_{3}}{12}\\
 & + & \left\{ \frac{337\pi^{2}}{54}-\frac{8707}{144}+\frac{107\zeta_{3}}{6}-\frac{7\pi^{4}}{30}+\left(\frac{3\pi^{2}}{4}-\frac{739}{48}\right)\ln\rho+\frac{13\ln^{2}\rho}{8}\right\} \rho\nonumber \\
 & + & \left\{ \frac{11\pi^{4}}{18}-\frac{1813}{8}-\frac{19\pi^{2}}{6}-\frac{685}{6}\ln\rho+4\ln^{2}\rho\right\} \rho^{2}\nonumber \\
 & + & \left\{ \frac{7309\pi^{2}}{36}-\frac{8039}{16}+\frac{47\pi^{4}}{30}-\frac{1136\pi^{2}\ln{2}}{3}-\frac{333\zeta_{3}}{2}+20\pi^{2}\zeta_{3}+120\zeta_{5}\right.\nonumber \\
 &  & \left.+\left(36\zeta_{3}-\frac{6867}{16}-\frac{541\pi^{2}}{12}+2\pi^{4}\right)\ln\rho+\left(3\pi^{2}-\frac{893}{8}\right)\ln^{2}\rho+\frac{47}{2}\ln^{3}\rho\right\} \rho^{3}\nonumber \\
 & + & \left\{ \frac{8947}{32}+\frac{5977\pi^{2}}{12}+\frac{2\pi^{4}}{45}-\frac{2272\pi^{2}\ln{2}}{3}-\frac{705\zeta_{3}}{2}+\frac{100\pi^{2}\zeta_{3}}{3}+200\zeta_{5}\right.\nonumber \\
 &  & \left.+\left(84\zeta_{3}-\frac{1049}{2}-\frac{171\pi^{2}}{2}+\frac{10\pi^{4}}{3}\right)\ln\rho+\left(7\pi^{2}-\frac{271}{2}\right)\ln^{2}\rho+16\ln^{3}\rho\right\} \rho^{4}\nonumber \\
 & + & \left\{ \frac{8313}{16}+\frac{518717\pi^{2}}{1800}-\frac{83\pi^{4}}{30}-\frac{1136\pi^{2}\ln{2}}{3}-\frac{459\zeta_{3}}{2}+20\pi^{2}\zeta_{3}+120\zeta_{5}\right.\nonumber \\
 &  & \left.+\left(72\zeta_{3}-\frac{4737}{16}-\frac{2293\pi^{2}}{60}+2\pi^{4}\right)\ln\rho+\left(6\pi^{2}-\frac{871}{8}\right)\ln^{2}\rho+\frac{33}{2}\ln^{3}\rho\right\} \rho^{5}\nonumber \end{eqnarray}
\begin{eqnarray}
V_{L} & = & -~\frac{1009}{288}+\frac{8\zeta_{3}}{3}+\frac{77\pi^{2}}{216}+\left\{ \frac{215}{18}-\frac{16\zeta_{3}}{3}-\frac{43\pi^{2}}{54}+\frac{13}{6}\ln\rho-\ln^{2}\rho\right\} \rho\label{eqn:vl}\\
 & + & \left\{ \frac{118}{3}-\frac{4\pi^{2}}{3}+\frac{52}{3}\ln\rho-8\ln^{2}\rho\right\} \rho^{2}\nonumber \\
 & + & \left\{ \frac{163}{2}+48\zeta_{3}-\frac{293\pi^{2}}{18}+\left(\frac{93}{2}+\frac{20\pi^{2}}{3}\right)\ln\rho-\ln^{2}\rho-20\ln^{3}\rho+\frac{64\pi^{2}\ln2}{3}\right\} \rho^{3}\nonumber \\
 & + & \left\{ 76\zeta_{3}-33-\frac{133\pi^{2}}{3}+\left(39+16\pi^{2}\right)\ln\rho+52\ln^{2}\rho-32\ln^{3}\rho+\frac{128\pi^{2}\ln2}{3}\right\} \rho^{4}\nonumber \\
 & + & \left\{ \frac{1}{2}+36\zeta_{3}-\frac{2417\pi^{2}}{90}+\frac{64\pi^{2}\ln2}{3}+\left(\frac{14\pi^{2}}{3}-\frac{195}{2}\right)\ln\rho+109\ln^{2}\rho-36\ln^{3}\rho\right\} \rho^{5},\nonumber \\
V_{H} & = & \frac{16987}{576}-\frac{64\zeta_{3}}{3}-\frac{85\pi^{2}}{216}+\left\{ \frac{35\pi^{2}}{54}-\frac{4109}{216}+\frac{32\zeta_{3}}{3}\right\} \rho+\left\{ \frac{8\pi^{2}}{3}-\frac{1198}{45}\right\} \rho^{2}\label{eqn:vh}\\
 & + & \left\{ \frac{9\pi^{2}}{2}-16\zeta_{3}-\frac{30043}{1080}+\left(\frac{20\pi^{2}}{3}-\frac{1193}{18}\right)\ln\rho\right\} \rho^{3}+\left\{ \frac{156901877}{2116800}-\frac{11\pi^{2}}{18}-64\zeta_{3}\right.\nonumber \\
 &  & \left.+\left(\frac{20\pi^{2}}{3}-\frac{186689}{2520}\right)\ln\rho\right\} \rho^{4}+\left\{ \frac{24416711}{352800}-24\zeta_{3}-\frac{9\pi^{2}}{2}+\left(6\pi^{2}-\frac{9883}{140}\right)\ln\rho\right\} \rho^{5},\nonumber 
\\
V_{C} & = & -~\frac{1009}{288}+\frac{8}{3}\zeta_{3}+\frac{77}{216}\pi^{2}+\left\{ \frac{94}{9}-\frac{16\zeta_{3}}{3}-\frac{167\pi^{2}}{108}+\frac{13}{6}\ln\rho-\ln^{2}\rho\right\} \rho\label{eqn:vc}\\
 & + & \left\{ \frac{145}{3}+\frac{22\pi^{2}}{3}+\frac{52}{3}\ln\rho-8\ln^{2}\rho\right\} \rho^{2}\nonumber \\
 & + & \left\{ 68+120\zeta_{3}-20\ln^{3}\rho+\frac{1061\pi^{2}}{36}+\left(\frac{129}{2}+\frac{106\pi^{2}}{3}\right)\ln\rho-\ln^{2}\rho\right\} \rho^{3}\nonumber \\
 & + & \left\{ 196\zeta_{3}-\frac{4483}{36}-\frac{121\pi^{2}}{18}+\left(\frac{599}{6}+\frac{158\pi^{2}}{3}\right)\ln\rho+44\ln^{2}\rho-32\ln^{3}\rho\right\} \rho^{4}\nonumber \\
 & + & \left\{ 200\zeta_{3}-\frac{20557}{72}-\frac{185\pi^{2}}{6}+\left(\frac{98\pi^{2}}{3}+\frac{686}{3}\right)\ln\rho-36\ln^{2}\rho-12\ln^{3}\rho\right\} \rho^{5}.~~~~~~~~~\nonumber \end{eqnarray}

\section{Master integrals}

\label{sec:mastersApp}
In this appendix, we collect new results for master integrals
that were obtained in this project. 

For the three-loop integrals needed for the factorized diagrams of
the type shown in Fig. \ref{fig:regs} (l), the $\order{\ep}$ term
of the integral $J_{3}$ of Ref.~\cite{Grozin:2006xm} was needed.
It was privately communicated to us by V. A. Smirnov,
\begin{equation}
\frac{J_{3}}{\pi^{2}\mathcal{F}^{3}}=-~\frac{32}{3}+\left(\frac{256\ln2}{3}-\frac{448}{3}-\frac{64\pi}{3}\right)\ep,\qquad\mathcal{F}\equiv\frac{\Gamma(1+\ep)}{(4\pi)^{D/2}}.\label{eq:u9}\end{equation}

For the un-factorized four-loop diagrams in Fig. \ref{fig:regs}(b),
the master integrals have already been classified and largely evaluated
in Ref.~\cite{vanRitbergen:1999fi}. Using the approach described
in Section \ref{sec:masters}, we obtained additional terms of their
expansion in $\ep$. To save space, we present here only those additional
terms, using the notation of Ref.~\cite{vanRitbergen:1999fi}, in
particular Eqs. (5.7-11) of that paper. Instead
of repeating here the long expressions already known, we define $\Delta_{N}$,
$N=A..F$, to denote the new terms. For any affected master integral,
its improved value can be found as follows, 
\begin{equation}
\mathrm{Im}\left[I_{N}\left(a_{1},\ldots,a_{11}\right)\right]\equiv\pi s^{j}\left(N_{\epsilon}\right)^{4}\left[\mathrm{terms\, already\, known}+\Delta_{N}\left(a_{1},\ldots,a_{11}\right)\right],
\label{eq:defDelta}
\end{equation}
where, as defined in Ref.~\cite{vanRitbergen:1999fi}, $N_{\epsilon}=\frac{\pi^{2}}{\left(\pi s\right)^{\epsilon}}\frac{\Gamma^{2}\left(1-\epsilon\right)\Gamma\left(1+\epsilon\right)}{\Gamma\left(1-2\epsilon\right)}$
and $s$ denotes the square of the momentum flowing into the diagram.
Its power $j$ depends on the integral considered and, for each one,
can be found in Ref.~\cite{vanRitbergen:1999fi}. Table \ref{tab:masters}
shows the extra terms $\Delta_{N}$. 

We use the following notation for transcendental constants, $A_{n}=\Li_{n}\frac{1}{2}$,
$n=4..6$, and $s_{6}=0.9874414...$. Here $\Li_{n}$ denotes polylogarithms
\cite{lewin}, $\Li_{n}x=\sum_{i=1}^{\infty}\frac{x^{i}}{i^{n}}$
and $s_{6}\equiv S_{-5,-1}\left(\infty\right)$ is certain harmonic
sum \cite{Vermaseren:1998uu,Maitre:2005uu}.

\begin{table}[h]
\caption{\label{tab:masters}Additional terms for master integrals of
Ref. \protect\cite{vanRitbergen:1999fi}, as defined in Eq.~(\ref{eq:defDelta}).}
\begin{ruledtabular}
\begin{tabular}{l@{\hspace{5mm}}l}
$\Delta_{A}(1,1,1,1,0,1,1,1,1,1,0)$  & $55\zeta_{5}-\frac{2\pi^{4}}{9}-\frac{13\pi^{2}\zeta_{3}}{2}+\ep\left(\frac{143\pi^{6}}{3240}-\frac{2\pi^{4}}{3}-26\pi^{2}\zeta_{3}-51\zeta_{3}^{2}+220\zeta_{5}\right)$ \tabularnewline
$\Delta_{A}(1,1,1,0,1,0,1,1,1,0,0)$  & $\ep^{2}\left(\frac{7945\pi^{2}}{48}-\frac{119727}{32}+\frac{269\pi^{4}}{72}+\frac{3185\zeta_{3}}{4}-\frac{49\pi^{2}\zeta_{3}}{3}+279\zeta_{5}\right)$ \tabularnewline
$\Delta_{A}(1,0,1,1,0,1,0,1,0,1,0)$  & $\ep^{3}\left(\frac{2515231}{1944}-\frac{101531\pi^{2}}{2592}-\frac{1309\pi^{4}}{180}-\frac{35951\zeta_{3}}{72}-\frac{7\pi^{2}\zeta_{3}}{3}-\frac{779\zeta_{5}}{2}\right)$ \tabularnewline
$\Delta_{A}(1,1,1,1,0,1,0,1,0,1,0)$  & $\ep^{2}\left(\frac{14437}{16}+\frac{169\pi^{2}}{16}+\frac{37\pi^{4}}{90}+\frac{5\zeta_{3}}{4}-41\zeta_{5}\right)$ \tabularnewline
$\Delta_{A}(1,0,1,0,1,1,1,0,1,0,0)$  & $\ep^{3}\left(\frac{51175105}{15552}-\frac{238819\pi^{2}}{2592}-\frac{3149\pi^{4}}{360}-\frac{54895\zeta_{3}}{72}+\frac{43\pi^{2}\zeta_{3}}{3}-\frac{1243\zeta_{5}}{2}\right)$ \tabularnewline
$\Delta_{A}(1,1,1,0,1,1,1,0,1,0,0)$  & $\ep^{2}\left(\frac{3587\pi^{2}}{48}-\frac{64177}{32}+\frac{367\pi^{4}}{180}+\frac{1245\zeta_{3}}{4}-\frac{4\pi^{2}\zeta_{3}}{3}+59\zeta_{5}\right)$ \tabularnewline
$\Delta_{A}(1,0,1,1,1,1,0,0,1,1,0)$  & $\ep^{2}\left(\frac{493\pi^{2}}{8}-\frac{119727}{32}+\frac{161\pi^{4}}{15}+600\zeta_{3}-\frac{20\pi^{2}\zeta_{3}}{3}+1092\zeta_{5}\right)$ \tabularnewline
$\Delta_{A}(1,0,1,1,0,1,1,1,1,0,0)$  & $\ep^{2}\left(\frac{3331\pi^{4}}{360}-\frac{147441}{64}-\frac{149\pi^{2}}{16}+\frac{1123\zeta_{3}}{4}+\frac{19\pi^{2}\zeta_{3}}{3}+759\zeta_{5}\right)$ \tabularnewline
$\Delta_{A}(1,1,1,0,0,1,1,1,1,0,0)$  & $\ep^{2}\left(\frac{4763\pi^{2}}{48}-\frac{38519}{16}+\frac{125\pi^{4}}{24}+\frac{2145\zeta_{3}}{4}-\frac{101\pi^{2}\zeta_{3}}{3}+563\zeta_{5}\right)$ \tabularnewline
$\Delta_{B}(1,1,1,1,1,1,1,1,1,1,1)$  & $\frac{1045\zeta_{5}}{32}-\frac{17\pi^{4}}{360}-\frac{77\pi^{2}\zeta_{3}}{32}$ \tabularnewline
$\Delta_{B}(1,1,1,1,1,1,0,1,1,1,0)$  & $-\frac{17\pi^{4}}{60}-\frac{79\pi^{2}\zeta_{3}}{48}-\frac{355\zeta_{5}}{16}$ \tabularnewline
$\Delta_{B}(1,0,1,0,1,1,1,0,1,1,0)$  & $\ep^{3}\left(72\pi^{2}-\frac{124763}{32}+\frac{793\pi^{4}}{60}+\frac{1961\zeta_{3}}{2}-\frac{43\pi^{2}\zeta_{3}}{3}+814\zeta_{5}\right)$ \tabularnewline
$\Delta_{B}(1,0,1,0,1,1,1,0,1,1,0)$  & $\ep\left(\frac{1391}{2}-\frac{58\pi^{2}}{3}-\frac{9\pi^{4}}{20}-68\zeta_{3}+\pi^{2}\zeta_{3}-27\zeta_{5}\right)$ \tabularnewline
$\Delta_{B}(1,1,1,0,1,0,1,1,1,1,0)$  & $\ep^{2}\left(2172-\frac{317\pi^{2}}{6}-\frac{401\pi^{4}}{90}-472\zeta_{3}+\frac{25\pi^{2}\zeta_{3}}{3}-232\zeta_{5}\right)$ \tabularnewline
$\Delta_{B}(1,1,1,0,1,1,1,0,1,0,0)$  & $\ep\left(24A_{4}+6A_{5}-\frac{\ln^{5}{2}}{20}-6\pi^{2}\ln{2}+\ln^{4}{2}-\frac{49\pi^{4}}{180}
 +\frac{19\pi^{4}\ln{2}}{80}+15\zeta_{3}-\frac{51\pi^{2}\zeta_{3}}{16}+\frac{299\zeta_{5}}{32}\right)$ \tabularnewline
$\Delta_{B}(1,0,1,1,1,1,0,0,2,2,0)$  & $\ep^{2}\left(\frac{\ln^{5}{2}}{20}-12A_{4}-6A_{5}-\frac{\ln^{4}{2}}{2}-\frac{33\pi^{2}}{2}-\pi^{2}\ln{2}-\frac{11\pi^{4}}{60}-\frac{19\pi^{4}\ln{2}}{80}
 - \frac{199\zeta_{3}}{2}+\frac{173\pi^{2}\zeta_{3}}{16}+\frac{5057\zeta_{5}}{32}\right)$
\tabularnewline
 & $+\ep^{3}\left(12A_{4}-12A_{5}-6A_{6}-\frac{159s_{6}}{2}
 +\frac{\ln^{4}{2}}{2}+\frac{\ln^{5}{2}}{10}-\frac{\ln^{6}{2}}{120}-\frac{131\pi^{2}}{2}+\frac{99A_{4}\pi^{2}}{2}-17\pi^{2}\ln{2}-\frac{31\pi^{4}}{4}\right.$ \tabularnewline
 & ~~$\left.+\frac{33\pi^{2}\ln^{4}{2}}{16}-\frac{975\zeta_{3}}{2}+\frac{47\pi^{2}\zeta_{3}}{3}-\frac{19\pi^{4}\ln{2}}{40}-\frac{311\pi^{4}\ln^{2}{2}}{160}+\frac{51\pi^{2}\zeta_{3}\ln{2}}{2}
 + \frac{7213\pi^{6}}{5040}+\frac{2177\zeta_{3}^{2}}{16}-\frac{1749\zeta_{5}}{16}\right)$ \tabularnewline
$\Delta_{C}(1,0,1,1,1,0,0,1,1,1,0)$  & $\ep\left(\frac{167\pi^{2}}{12}-\frac{10325}{32}-4\pi^{2}\ln{2}+\frac{149\pi^{4}}{360}+105\zeta_{3}\right)$ \tabularnewline
$\Delta_{C}(1,0,1,1,1,1,0,1,1,1,0)$  & $\ep\left(113-48A_{4}-2\ln^{4}{2}+\frac{151\pi^{2}}{6}+4\pi^{2}\ln{2}-\frac{13\pi^{4}}{60}+14\zeta_{3}\right)$ \tabularnewline
 & $+\ep^{2}\left(934-576A_{4}-480A_{5}-24\ln^{4}{2}+4\ln^{5}{2}+\frac{356\pi^{2}}{3}+42\pi^{2}\ln{2}-8\pi^{2}\ln^{2}{2}+\frac{8\pi^{2}\ln^{3}{2}}{3}\right.$ \tabularnewline
 & ~~$\left.+\frac{5\pi^{4}}{72}-\frac{3\pi^{4}\ln{2}}{2}+24\zeta_{3}-\frac{59\pi^{2}\zeta_{3}}{6}+\frac{913\zeta_{5}}{4}\right)$ \tabularnewline
$\Delta_{C}(1,0,1,1,1,1,0,1,1,1,0)$  & $\ep^{2}\left(\frac{131899}{64}-92\pi^{2}\ln{2}+192A_{4}+8\ln^{4}{2}+\frac{543\pi^{2}}{16}+16\pi^{2}\ln^{2}{2}
 -\frac{99\pi^{4}}{40}+\frac{831\zeta_{3}}{4}+\pi^{2}\zeta_{3}-45\zeta_{5}\right)$ \tabularnewline
$\Delta_{E}(1,1,1,0,1,1,1,1,1,1,0)$  & $\frac{13\pi^{2}\zeta_{3}}{3}-\frac{4\pi^{4}}{15}-94\zeta_{5}+\ep\left(\frac{52\pi^{2}\zeta_{3}}{3}-\frac{4\pi^{4}}{5}-\frac{217\pi^{6}}{1620}+6\zeta_{3}^{2}-376\zeta_{5}\right)$ \tabularnewline
$\Delta_{F}(1,1,1,1,1,1,1,1,1,1,1)$  & $\frac{\pi^{4}}{30}+\frac{221\pi^{2}\zeta_{3}}{48}-\frac{799\zeta_{5}}{16}$ \tabularnewline
$\Delta_{F}(1,1,1,1,1,1,1,0,1,1,0)$  & $\frac{61\pi^{2}\zeta_{3}}{48}-\frac{37\pi^{4}}{180}-\frac{1019\zeta_{5}}{16}$ \tabularnewline
$\Delta_{F}(1,0,1,1,1,1,1,0,1,2,0)$  & $-~\ep\left(\frac{3\pi^{4}}{20}+\frac{107\pi^{2}\zeta_{3}}{48}+\frac{547\zeta_{5}}{16}\right)$ \tabularnewline
$\Delta_{F}(1,1,1,1,0,1,1,0,1,1,0)$  & $\ep\left(1031-\frac{199\pi^{2}}{6}-\frac{209\pi^{4}}{180}-163\zeta_{3}+\frac{7\pi^{2}\zeta_{3}}{3}-64\zeta_{5}\right)$ \tabularnewline
$\Delta_{F}(1,1,1,1,0,0,1,1,1,1,0)$  & $\ep^{2}\left(\frac{253\pi^{2}}{2}-2172+\frac{407\pi^{4}}{90}+899\zeta_{3}-\frac{98\pi^{2}\zeta_{3}}{3}+403\zeta_{5}\right)$ \tabularnewline
 & $+\ep^{3}\left(\frac{2007\pi^{2}}{2}+\frac{547\pi^{4}}{18}-17037+\frac{487\pi^{6}}{1134}
 +5869\zeta_{3}-396\zeta_{3}^{2}+4433\zeta_{5}-\frac{1078\pi^{2}\zeta_{3}}{3}\right)$ \tabularnewline
$\Delta_{F}(1,1,1,1,1,1,0,0,1,1,0)$  & $46A_{4}-30+\frac{23\ln^{4}{2}}{12}+\frac{7\pi^{2}}{6}-2\pi^{2}\ln{2}-\frac{5\pi^{2}\ln^{2}{2}}{3}-\frac{\pi^{4}}{9}+24\zeta_{3}
 +\ep\left(184A_{4}-272+46A_{5}+\frac{23\ln^{4}{2}}{3}
 \right.$\tabularnewline
 & ~~$\left.-\frac{23\ln^{5}{2}}{60}+\frac{67\pi^{2}}{6}-3\pi^{2}\ln{2}
 -\frac{73\pi^{4}}{360}-\frac{20\pi^{2}\ln^{2}{2}}{3}+\frac{5\pi^{2}\ln^{3}{2}}{9}-\frac{547\pi^{4}\ln{2}}{360}+\frac{209\zeta_{3}}{2}-\frac{99\pi^{2}\zeta_{3}}{8}+\frac{3921\zeta_{5}}{16}\right)$ \tabularnewline
 & \tabularnewline
\end{tabular}
\end{ruledtabular}
\end{table}


\end{document}